\documentclass[runningheads]{llncs}

\usepackage{graphicx}
\usepackage{amsmath,amssymb}
\usepackage{booktabs}
\usepackage{url}
\usepackage{hyperref}
\usepackage{comment}
\usepackage{upquote}
\title{Evaluating Post-Quantum and Hybrid Signatures for Blockchains: Performance, Storage and Ledger Growth}
\titlerunning{Post-Quantum and Hybrid Signatures for Blockchains}

   \author{Anthony Mikhaiel \and Essam Ghadafi}
   \authorrunning{A. Mikhaiel and E. Ghadafi}
   \institute{Newcastle University, Newcastle upon Tyne, UK\\
\email{A.Mikhaiel2@newcastle.ac.uk}\\
\email{Essam.Ghadafi@newcastle.ac.uk}
   }
\begin{document}

\maketitle

\begin{abstract}
Advances in quantum computing pose a long-term threat to blockchain systems
that rely on classical digital signature schemes such as ECDSA. Migrating these
systems to post-quantum cryptography introduces practical costs beyond
cryptographic security, including larger transactions, higher computational
costs and faster ledger growth. We present an empirical evaluation of 13
signature configurations in a common simulated blockchain framework: ECDSA,
ML-DSA-44/65/87, Falcon-512/1024, SLH-DSA-SHA2-128s, and six hybrids combining
ECDSA with each post-quantum scheme. We measure signing and verification costs,
transaction-processing throughput, and signature and public-key sizes, and
project the measured overheads onto five public blockchains: Bitcoin, Ethereum,
Solana, BNB Smart Chain and TRON. Verification remains inexpensive for the
evaluated lattice-based schemes, with Falcon-512 achieving a verification time
comparable to ECDSA (30.1 versus 32.8~$\mu$s). Storage overheads, however,
are substantial. Under our modelling assumptions, had the recorded transaction
histories used Falcon-512, the most compact post-quantum scheme evaluated,
ledger sizes would have been 79--229\% larger, depending on the blockchain.
Hybrid configurations introduce relatively modest additional storage overhead
compared with their standalone post-quantum counterparts, while
ECDSA+Falcon-512 approximately doubles the verification cost of standalone
Falcon-512. Existing transaction- and script-size limits also present obstacles
to straightforward migration in Bitcoin and Solana. Overall, our findings
suggest that storage overhead may pose a greater challenge than signature
verification costs when migrating blockchain systems to post-quantum
cryptography.
\end{abstract}


\section{Introduction}
\label{sec:intro}
Blockchain systems rely extensively on public-key cryptography to
establish ownership, authenticate transactions, and protect the integrity
of distributed ledgers. In many widely deployed blockchains, these
functions depend on digital signature schemes based on elliptic-curve
cryptography (ECC). The security of such schemes relies on the
computational hardness of the elliptic-curve discrete logarithm problem.
However, Shor's quantum algorithm can solve discrete logarithms in
polynomial time on a sufficiently powerful quantum computer~\cite{Shor94},
undermining the security assumptions on which ECC-based signatures depend.
Consequently, the development of cryptographically relevant quantum
computers would pose a fundamental threat to blockchains whose transaction
authentication continues to rely on schemes such as the Elliptic Curve
Digital Signature Algorithm (ECDSA). This threat is particularly concerning
because public keys may be exposed on-chain during transaction processing,
potentially leaving funds associated with those keys vulnerable to future
quantum attacks.

Recent resource estimates indicate that the 256-bit elliptic-curve
discrete logarithm problem over secp256k1, the curve used by Bitcoin and
Ethereum for transaction authentication, could be solved with at most
1,200 logical qubits and 90 million Toffoli gates. Under certain hardware
assumptions, such an attack could run in minutes on a superconducting
quantum computer using fewer than half a million physical
qubits~\cite{babbush2026securing}. Such capabilities could enable attacks
on transactions awaiting confirmation in some blockchains, in addition
to attacks on public keys already exposed on-chain.

 Migrating a public blockchain requires both protocol changes agreed upon by a decentralised community and action by individual users to secure their funds, a process that may take years. The costs of migration therefore need to be understood well before the threat materialises.

The transition to post-quantum cryptography (PQC) has therefore become
an increasingly important security objective. In 2024, the U.S.
National Institute of Standards and Technology (NIST) standardised
the lattice-based ML-DSA~\cite{FIPS204} and hash-based
SLH-DSA~\cite{fips205} digital signature schemes.
 The lattice-based
Falcon~\cite{falcon} was also selected for standardisation as FN-DSA
(FIPS 206, currently in development)~\cite{nistir8413}.
 NIST has also set out a transition timeline under which quantum-vulnerable
digital signatures, including ECDSA, are deprecated after 2030 and disallowed
after 2035~\cite{nistir8547}.
Governments and security agencies have likewise begun publishing
roadmaps for migration away from quantum-vulnerable public-key
cryptography~\cite{EUPQCStrategy24,NCSCMigration}.
For blockchain systems, however, replacing a classical signature
scheme with a post-quantum alternative is not simply a cryptographic
substitution. Signatures and public keys may form part of the data processed, communicated, verified, and stored by blockchains. Differences in signature size, public-key size, and computational cost can therefore affect not only individual cryptographic operations but also transaction size, processing performance, throughput, resource consumption, block utilisation, and ultimately ledger growth.

These considerations are particularly important because blockchain data are replicated and retained across participating nodes. An increase in the size of individual transactions may appear modest when considered in isolation, but can translate into substantial cumulative storage and communication overhead as transaction volumes and ledger lifetimes 
increase. Similarly, differences in signature verification cost can become significant when nodes must validate large numbers of transactions. Migration strategies that combine classical and post-quantum signatures may provide a transitional approach in which both forms of authentication are retained, but this additional protection comes with further computational and storage costs. Understanding these system-level consequences is therefore necessary when assessing the practicality of post-quantum migration for blockchain systems.

 This paper investigates these trade-offs through a
controlled experimental evaluation of classical,
post-quantum, and hybrid digital signatures in a
simulated blockchain environment. The evaluation
considers ECDSA, multiple parameter sets of ML-DSA
and Falcon, SLH-DSA-SHA2-128s, and hybrid
configurations combining ECDSA with each
post-quantum scheme. Using a common transaction
workload, we examine how signature choice and
security level affect cryptographic performance,
transaction processing, throughput, and storage
requirements. We further use the measured storage
overheads to estimate their potential implications
for five blockchain systems under explicit migration assumptions.

The main contributions of this work are as follows:
\begin{itemize}
    \item We present a controlled experimental comparison of classical, post-quantum, and hybrid digital signature schemes in a common simulated blockchain environment, covering multiple parameter sets of ML-DSA and Falcon, as well as SLH-DSA.

    \item We evaluate the computational and storage
    implications of these configurations, examining
    signing and verification costs alongside
    transaction processing, throughput, transaction
    size, and ledger growth.

    \item We estimate the potential storage
    implications of post-quantum signature migration
    for five public blockchain systems using
    measured signature overheads and observed
    transaction activity, under clearly stated
    assumptions.
\end{itemize}

The remainder of this paper is organised as follows. Section~\ref{sec:related} reviews related work. Section~\ref{sec:experiment} describes the experimental framework, signature configurations, workloads, and evaluation methodology. Section~\ref{sec:results} presents the experimental results. Section~\ref{sec:discussion} discusses the resulting trade-offs and their implications for post-quantum blockchain migration. Finally, Section~\ref{sec:summary} concludes the paper and identifies directions for future work.

\section{Related Work}
\label{sec:related}
Several studies have examined the cost of replacing classical blockchain signatures with post-quantum alternatives. Marchsreiter~\cite{Marchsreiter2025} 
implemented a proof-of-work blockchain and compared ECDSA with several post-quantum signature schemes, including Dilithium, Falcon, SPHINCS+, and XMSS. Alongside cryptographic timings, the study measured transaction and block sizes and throughput on desktop and embedded hardware. It also implemented and evaluated public-key recovery for Falcon to reduce transaction sizes. The evaluation focuses on the implemented blockchain prototype and does not examine classical--post-quantum hybrid configurations or project the measured overheads onto existing public blockchains.
Thanalakshmi et al.~\cite{Thanalakshmi2023} 
compared Dilithium, Falcon
and SPHINCS+ with ECDSA in a UTXO-based proof-of-work prototype,
measuring transaction sizes and block mining times. To reduce on-chain
storage, they recorded only hashes of signatures and public keys
on-chain and stored the full values in the InterPlanetary File System
(IPFS). While this approach reduces on-chain storage requirements,
it introduces a dependency on the availability of off-chain data.
Our projections instead assume that signatures remain on-chain.

Holmes~\cite{holmes2023impact} used historical Bitcoin
statistics to estimate the impact of post-quantum signatures
on transaction sizes, block sizes and the number of
transactions per block, finding a reduction in the latter
by a factor of 2 to 32 for a fixed block size. The study
highlighted the substantial storage overheads associated with
migrating from ECDSA to post-quantum signatures.
Zakharov et al.~\cite{zakharov2026lattice} examined Dilithium, Falcon and
Hawk for Bitcoin's post-quantum transition, considering security, combined
public-key and signature sizes, and deployment constraints. They identified
Falcon as the most promising lattice-based candidate despite its
implementation complexity.
Mallick et al.~\cite{mallick2025quantum}
systematised the quantum vulnerabilities of seven layer-1 blockchains and
analytically examined the impact of replacing classical signatures with
ML-DSA-44, Falcon-512 and SLH-DSA-128f on theoretical throughput,
consensus-signature aggregation and ledger storage. Their storage analysis
relied on worst-case bounds and assumed proportions of ledger data
attributable to signatures and public keys. Hybrid configurations were
discussed but not experimentally evaluated.

Other work has focused more directly on execution costs. Juaristi et al.~\cite{Juaristi2024}  deployed a private Go-Ethereum network and used transaction hashes from the network to benchmark ECDSA against Dilithium,
Falcon-1024, and SPHINCS+, measuring execution time, memory usage, and
key and signature sizes. The benchmarks ran in a separate backend rather than replacing the nodes' native ECDSA transaction validation. 
Schemitt et al.~\cite{Schemitt2025} benchmarked a broad range of post-quantum
signature implementations and used the measured verification times to simulate
block verification in Bitcoin- and Ethereum-like environments. Their study
connects cryptographic benchmarks to block processing but does not quantify
the impact of larger signatures and public keys on transaction sizes or ledger
growth. They report faster verification for ML-DSA than ECDSA across their
evaluated security levels. However, their ECDSA baseline uses the NIST curves
P-256, P-384, and P-521 through a general-purpose cryptographic library,
whereas we benchmark ECDSA over secp256k1, the curve used by Bitcoin,
Ethereum, BNB Smart Chain, and TRON, using libsecp256k1. In our experiments,
ECDSA verification is faster than ML-DSA verification, although our ML-DSA
measurements do not use a dedicated AVX2 implementation (Section~\ref{sec:implement}).

Hybrid migration has also received attention. Holcomb et
al.~\cite{Holcomb2020} developed PQFabric, integrating hybrid classical
and post-quantum signatures into Hyperledger Fabric's identity management
and transaction authentication mechanisms, with support for crypto-agility
and gradual migration. Their evaluation shows that larger signatures
and public keys can increase hashing costs, sometimes making them
comparable to the time spent on signing and verification. In an Ethereum setting, Ahmed and Sisinni~\cite{AhmedSisinni2026}
combined native ECDSA transaction authentication with additional
post-quantum signatures embedded in transaction payloads and verified
off-chain. They measured signing and verification times, public-key
and signature sizes, and gas costs for key registration and transactions.
Since transaction validity continues to depend on ECDSA, the
post-quantum signatures are not enforced by Ethereum's native
transaction-validation rules. Rohit  and Sridhar ~\cite{kc2026security} evaluated
a parallel dual-signature construction combining Falcon-512
and ML-DSA-65 for blockchain transaction authentication,
measuring computational performance, signature sizes and
resource usage in a lightweight Python model. They also
demonstrated resistance to cross-context replay in their
experimental setting. Unlike our hybrid configurations,
their construction combined two lattice-based post-quantum
signature schemes.
Campbell~\cite{Campbell2026Hybrid} examined the challenges of
migrating Bitcoin and Ethereum to hybrid classical and
post-quantum signatures, considering relevant BIP and EIP
proposals, performance overheads, storage growth, and
governance constraints. The study draws partly on performance measurements from
permissioned blockchain systems to estimate the potential impact on
Bitcoin and Ethereum, while acknowledging the uncertainty of such extrapolations.
Lie and Marsh~\cite{lie2026domain} examined blockchain-specific
requirements for post-quantum signatures, including canonical
encodings, worst-case verification costs, hybrid migration
and signature aggregation. They argued that existing
post-quantum signature schemes are not sufficient as
drop-in replacements for the signature layer of modern
public blockchains.

There are also proposals to reduce the cost of verifying post-quantum signatures within Ethereum. The draft EIP-7619~\cite{EIP7619} specifies a Falcon-512 verification precompile and proposes a gas schedule based on verification benchmarks. Kysil et al.~\cite{Kysil2025} implemented and evaluated
post-quantum signature verification in the Ethereum Virtual
Machine, considering W-OTS+, XMSS, SPHINCS+, and MAYO.
They compared full on-chain verification with an optimistic
approach based on Naysayer proofs, focusing primarily on
verification gas costs. Although they discuss storage costs
and provide rough lower-bound estimates of Ethereum's
post-quantum migration costs, they do not systematically
evaluate the impact on transaction processing performance
or long-term ledger growth.

The storage overhead of post-quantum authentication has also
motivated approaches that reduce repeated storage of large
cryptographic objects. Wang et al.~\cite{Wang2026} proposed a
Dilithium-Prime-based UTXO design in which public keys are
registered and subsequently referenced rather than repeatedly
included in transactions, although public-key reuse can increase
transaction linkability. They also investigated private-key
storage optimisation by regenerating private keys from stored
seeds. Finlow-Bates et al.~\cite{FinlowBates2026} examined the
infrastructure costs of post-quantum migration in replicated
blockchain networks, using simplified models to estimate storage
requirements and proposing a hash-based commit--reveal
alternative. Their analysis illustrates how increases in
transaction size can translate into substantial network-wide
costs, although their projections are based on modelling
assumptions rather than comparative benchmarks of multiple
signature schemes.

Algorand illustrates an incremental approach to post-quantum
migration. Its v5.0.0 upgrade (August 2026) introduced native
support for accounts based on a deterministic variant of
Falcon-1024, alongside existing Ed25519
accounts~\cite{algorand2026pqaccounts,algorand2026upgrade}.
Native Falcon-1024 transactions carry both the public key and the signature and, at the time of writing, incur three times the minimum fee for Ed25519 transactions, which
Algorand attributes to the additional block space they
consume~\cite{algorand2026pqaccounts,algorand2026upgrade}.
This is consistent with the storage overheads examined in
this paper. However, Algorand's consensus mechanism still
relies, at least in part, on classical
cryptography~\cite{algorand2026pqledger}.

Taken together, these studies and deployments show that post-quantum migration involves more than the computational cost of signing and
verification. They also highlight how overheads depend on
blockchain architecture, transaction representation, and
integration strategy. 
In this context, our study compares 13 classical, post-quantum, and hybrid signature
configurations in a common simulated blockchain environment:
ECDSA, three ML-DSA parameter sets, two Falcon parameter sets,
SLH-DSA-SHA2-128s, and six corresponding ECDSA hybrids. We
evaluate processing and storage overheads and use observed
transaction activity from five public blockchains to estimate
signature-related storage and ledger growth under explicitly
stated migration assumptions. These projections are illustrative
and do not represent deployments of the evaluated schemes on the
production networks.

\section{Methodology and Experimental Setup}
\label{sec:experiment}
Our study evaluates 13 classical, post-quantum and hybrid digital signature configurations within a common simulated blockchain environment. The signature configuration is the primary experimental variable, while transaction generation, verification and block-processing procedures are identical across all configurations. Each configuration is evaluated under two verification modes: verification immediately after signing and verification during block commitment. We measure both the direct computational costs of cryptographic operations and their effects on transaction processing and storage in the prototype, and then project the measured overheads onto five public blockchains.

\subsection{Blockchain Model and Signature Configurations}
The experiments use a local blockchain simulation that supports transaction
creation and signing, block formation and commitment, and subsequent chain
validation. Transactions are grouped into blocks and appended to a
hash-linked chain starting from a genesis block. The simulation does not
model consensus, peer-to-peer networking, mempool behaviour, or network
propagation. These components are outside the scope of this study, which
focuses specifically on the effects introduced by changing the digital
signature scheme.
We evaluate 13 signature configurations: ECDSA over secp256k1 with
SHA-256; ML-DSA-44, ML-DSA-65, ML-DSA-87, Falcon-512, Falcon-1024,
and SLH-DSA-SHA2-128s; and six hybrid configurations combining
ECDSA with each of the six aforementioned post-quantum schemes.
In a hybrid configuration, the classical and post-quantum schemes
independently sign the same transaction data, and both signatures are
stored and verified. Both signatures must verify successfully for a hybrid
transaction to be accepted. If either verification fails, the transaction
is rejected.

Each transaction contains a sender identifier, nonce, payload, signature(s),
the corresponding public key(s), algorithm identifier(s), and a transaction
identifier. The input to the signature scheme consists of the sender
identifier, nonce, and SHA-256 hash of the payload. The transaction identifier
is obtained by hashing the same canonical unsigned transaction representation.
Six simulated senders are used, with each sender maintaining its own nonce
sequence and key pairs. Payload sizes rotate through 16, 32, 64, 128, and
256 bytes. Key pairs are generated before timing begins and are retained in
memory throughout each run. This means that key-generation time is not
included in the performance measurements.
\subsection{Workload and Experimental Design}
\label{sec:workload}
Each run processes 10,000 transactions, grouped into blocks of 1,000
transactions, giving ten blocks per run. The same workload is used for all
13 signature configurations, so per-transaction timings, throughput and
ledger totals are directly comparable across configurations. Per-transaction
values are obtained by dividing cumulative times by the number of transactions
processed.    For the five configurations with the fastest signing, an additional
   100,000-transaction workload yielded per-transaction signing and verification
   times within 1.5\% of those reported here.

All other experimental parameters are kept fixed. The same six senders,
payload-size sequence, transaction format, signing input and block-processing
logic are used for each signature configuration, and payload generation is
deterministic.

Two verification modes are considered. In \emph{scheme-level verification}, a
transaction is verified immediately after it is signed. In \emph{commit-level
verification}, signature verification takes place when the block containing the
transaction is committed. The signature implementations are unchanged between
the two modes; only the point at which verification is performed differs.
Results for the two modes are therefore analysed separately.

Each configuration is run 20 times: ten runs with scheme-level verification and
ten with commit-level verification. 
For timing and throughput measurements, we report the median across the ten
runs of each verification mode, as it is less sensitive than the mean to
occasional outlier runs. Signature sizes, which vary for ECDSA and Falcon, are reported as means over
all generated signatures.
\subsection{Implementation and Experimental Environment}
\label{sec:implement}
The experiments were run on an ASUS Zenbook Pro 14 OLED with an Intel Core
i9-13900H processor and 32~GB of RAM, running Windows 11 and Python 3.13.
ECDSA is implemented using \texttt{coincurve} 21.0.0. The post-quantum
schemes are implemented using \texttt{liboqs-python} 0.16.0 with
\texttt{liboqs} 0.16.0, and system measurements are collected using
\texttt{psutil} 7.2.2. The liboqs build for this project uses the dedicated AVX2 implementations
of Falcon-512 and Falcon-1024, while ML-DSA and SLH-DSA use their C
implementations, which may also execute AVX2 instructions. The dedicated AVX2 implementation of ML-DSA was not
available in our Windows build of \texttt{liboqs} 0.16.0. The library was built with \texttt{GCC} 13.2.0, and the build
log and disassembly confirm that the dedicated Falcon AVX2 routines
were used for signing and verification.
 During the experiments, the machine was connected to
mains power, cooling was set to maximum capacity, and background applications
were closed.

Falcon is evaluated as implemented in \texttt{liboqs} 0.16.0; the forthcoming
FN-DSA standard may differ from this implementation, and the reported
measurements should not be interpreted as benchmarks of FN-DSA.

Before each run, the payloads and required key pairs are prepared and the
blockchain data structures are initialised. These setup operations are
excluded from the timed measurements. Timing starts immediately before the
transaction-processing loop.

For each transaction, the unsigned transaction data are constructed and
signed using the selected signature configuration. In scheme-level
verification mode, the signature is also verified at this point. The
transaction is then added to the current block. Once 1,000 transactions have
been collected, the block is committed. In commit-level verification mode,
signature verification takes place during this operation. Timing stops after
all transactions have been processed and all blocks have been committed.


\subsection{Metrics and Calculations}
\label{sec:metrics}
\paragraph{Cryptographic metrics.}
We measure signing and verification times separately. Under
scheme-level verification, total cryptographic time is
\[
    T_{\mathrm{crypto}} = T_{\mathrm{sign}} + T_{\mathrm{verify}},
\]
where both terms are cumulative times over the $N$ transactions
processed in a run. Crypto-only throughput is calculated as $\frac{N}{T_{\mathrm{crypto}}}$, with time expressed in seconds.

\paragraph{System-level metrics.}
End-to-end throughput of the simulated blockchain is calculated as
$\frac{N}{T_{\mathrm{elapsed}}}$, where $T_{\mathrm{elapsed}}$ is the duration of the timed
transaction-processing section, including block commitment.
Block commitment time is the cumulative time spent committing
blocks during a run. When reported in $\mu$s/tx, it is divided
by $N$ and converted to microseconds. In commit-level verification
mode, this measurement includes signature verification performed
during commitment.
\begin{table}[t]
\centering
\caption{Public-key and signature sizes (in bytes) of the evaluated schemes.
For variable-length signatures, the mean over all generated signatures is
reported, with the maximum encoded size for the evaluated implementation in
parentheses.}
\label{tab:sizes}
\begin{tabular}{|l|c|c|c|}
\hline
Scheme & NIST category & Public key  & Signature \\
\hline
ECDSA (secp256k1)   & --- &    33 & 70.49$^{a}$ (71) \\
ML-DSA-44           & 2   & 1,312 & 2,420 \\
ML-DSA-65           & 3   & 1,952 & 3,309 \\
ML-DSA-87           & 5   & 2,592 & 4,627 \\
Falcon-512          & 1   &   897 & 655.05$^{b}$ (752) \\
Falcon-1024         & 5   & 1,793 & 1,270.67$^{b}$ (1,462) \\
SLH-DSA-SHA2-128s   & 1   &    32 & 7,856 \\
\hline
\end{tabular}
\smallskip
\begin{minipage}{0.85\linewidth}
\raggedright
\footnotesize
$^{a}$ DER encoding, as produced by coincurve; length varies with the encoded
values of $r$ and $s$ (64~bytes in raw $r \| s$ form). Our projections use
the measured DER-encoded signature sizes for ECDSA and its hybrid
configurations, while existing-ledger baselines use each blockchain's
representative native signature size (Section~\ref{sec:realworld}).\\
$^{b}$ Compressed, unpadded encoding; length varies with the randomised
signing output.
Hybrid configurations store both public keys and both signatures, so their
sizes are the sums of their components.
\end{minipage}
\end{table}

\paragraph{Storage metrics.} Transaction and ledger sizes are measured from the
JSON-serialised representation used by the prototype. 
Public keys are stored as hexadecimal strings, with two JSON bytes per raw key
byte. We therefore report adjusted sizes that exclude the public keys
entirely, since public-key sizes are reported separately (Tables~\ref{tab:sizes} and \ref{tab:size-results}):
\[
S_{\mathrm{adj}} = S_{\mathrm{JSON}} - 2K, \qquad
L_{\mathrm{adj}} = L_{\mathrm{JSON}} - 2KN,
\]
where $S_{\mathrm{JSON}}$ and $L_{\mathrm{JSON}}$ are the serialised transaction
and ledger sizes, $K$ is the total raw public-key size in bytes per transaction,
and $N$ is the number of transactions. JSON field names, structural overhead and the payload are retained. The signature remains hex-encoded in the adjusted sizes, which are
therefore prototype-specific rather than native transaction sizes. Adjusted
transaction sizes are averaged over all transactions in a run.

\paragraph{Timing.} Individual signing, verification and block-commit operations
are timed with \texttt{time.perf\_counter\_ns()}, and the total elapsed time of
the timed section with \texttt{time.perf\_counter()}.
\subsection{Correctness and Validation}
\label{sec:validate}
After every run, the complete simulated blockchain is validated
by verifying all committed transaction signatures and checking
block hash-link integrity from the genesis block onwards.
The transaction workload is deterministic for a fixed set of
benchmark parameters. Signature sizes and timing results are
also checked for plausibility against expected library behaviour
and published values, recognising that execution times vary
across hardware platforms and implementations.

\subsection{Real-World 
Blockchain Data and Scaling}
\label{sec:realworld}
Public-ledger data are used to scale the benchmark results to real blockchain activity. 
Ledger sizes correspond to 20 August 2026, except for Solana, whose ledger size is approximate. Transaction statistics were collected in August 2026; those for BNB Smart Chain were computed from BscScan daily data up to 20 August 2026. 90-day averages are used where stated.

\begin{itemize}
\item Blockchains: Bitcoin, Ethereum, Solana, BNB Smart Chain and TRON.

\item Collected quantities: lifetime transactions, 90-day average daily
transactions, 90-day average transactions per block, 90-day average block
size (not available for Solana) and total ledger size. Storage uses decimal
megabytes (1~MB $= 10^6$ bytes).

\item Ledger size is defined as persistent historical block and transaction
data where this can be isolated. For Ethereum, the Geth full-node size is
used, which also includes the current (pruned) state, so baselines differ
in scope across chains.

\item For BNB Smart Chain, no suitable published ledger-size series was
identified, so we estimate cumulative block data as
$L_{\mathrm{BNB}} = \sum_d \bar{b}_d\, n_d$, where $\bar{b}_d$ and $n_d$
are BscScan's daily average block size (bytes) and daily block count,
summed from 29 August 2020 to 20 August 2026 inclusive. The estimate
reflects the block-size metric reported by BscScan and excludes node state.

\item Solana's ledger-size estimate is approximate. Its transaction counts
include validator vote transactions, with lifetime and daily counts using
the same definition. Since the Alpenglow upgrade of 21 October 2026
\cite{alpenglow} removed vote transactions from blocks, the Solana annual
projections, which use the pre-upgrade rate, are likely to overestimate
future signature storage and processing requirements.

\item Representative native signature sizes: Bitcoin 71-byte ECDSA;
Ethereum, BNB Smart Chain and TRON 65-byte ECDSA; Solana 64-byte Ed25519.
The estimated total signature size is lifetime transactions multiplied
by this size, a modelling estimate rather than a measurement of actual
historical signature bytes.

\item For Solana, the ECDSA-based hybrids are hypothetical. Based on
signature sizes alone, an Ed25519-based hybrid would require approximately
6.5 fewer bytes per transaction, a negligible difference for the
projected signature-storage overheads.
\end{itemize}

The projected ledger size for each configuration is calculated as
$L_c = L + N(s_c - s_{\mathrm{native}})$, keeping all other ledger data
unchanged, where $L$ is the existing ledger size, $N$ the number of lifetime
transactions, and $s_c$ and $s_{\mathrm{native}}$ the configuration's and the
chain's representative native signature sizes, respectively. These historical
projections are counterfactual: they estimate the size the recorded
transaction history would have occupied under each configuration, rather than
storage added by migration, since existing blocks would not be altered.

The annual projections instead estimate signature storage for future
transactions, extrapolating the 90-day average daily transaction count over
365 days at a constant rate. The projected additional storage for future
transactions is the difference between a configuration and the corresponding
native-signature baseline.

\begin{table}[t]
\centering
\small
\setlength{\tabcolsep}{4pt}
\caption{Public-ledger activity data (90-day averages where stated).}
\label{tab:chain-activity}
\begin{tabular}{|l|r|r|r|}
\hline
Blockchain & Lifetime TXs & \shortstack{90-day avg\\TX/block} & \shortstack{90-day avg\\daily TXs} \\
\hline
Bitcoin         & 1,411,916,537   & 4,596.73 & 668,294.17 \\
Ethereum        & 3,665,337,114   & 312.97   & 2,245,924.24 \\
Solana          & 537,114,837,352 & 1,343.90 & 283,892,018.53 \\
BNB Smart Chain & 14,329,656,901  & 82.01    & 15,734,662.34 \\
TRON            & 15,098,969,733  & 421.13   & 12,126,845.67 \\
\hline
\end{tabular}
\end{table}


\begin{table}[t]
\centering
\small
\setlength{\tabcolsep}{4pt}
\caption{Public-ledger storage and estimated signature data. Ledger sizes are as of 20 August 2026, except for Solana (approximate); 1~MB $= 10^6$ bytes.}
\label{tab:chain-storage}
\begin{tabular}{|l|r|r|r|}
\hline
Blockchain & \shortstack{90-day avg\\block size (B)} & Ledger size (MB) & \shortstack{Estimated signa-\\ture total (MB)} \\
\hline
Bitcoin         & 1,590,516 & 762,722.25                  & 100,246.07 (71 B/tx) \\
Ethereum        & 180,963.61 & 1,744,780.03$^{a}$         & 238,246.91 (65 B/tx) \\
Solana          & n/a        & $\sim$400,000,000$^{b}$    & 34,375,349.59 (64 B/tx) \\
BNB Smart Chain & 54,748.88  & 7,015,936.89$^{c}$         & 931,427.70 (65 B/tx) \\
TRON            & 100,471.46 & 3,896,903.48               & 981,433.03 (65 B/tx) \\
\hline
\end{tabular}
\smallskip
\parbox{\linewidth}{\raggedright\footnotesize
$^{a}$~Geth full-node (default sync) data size.~\\
$^{b}$~Approximate; no official measurement is published. An alternative archive estimate of approximately 505~TB is reported by Old Faithful~\cite{oldfaithful}.~\\
$^{c}$~Cumulative block data computed from BscScan daily average block size and daily block count; excludes node state.}
\end{table}

\paragraph{Data sources.}
Transaction statistics, block sizes and ledger sizes were obtained from
Blockchain.com~\cite{src-btc} for Bitcoin, Etherscan~\cite{src-eth,src-ethsync}
for Ethereum, the Solana Explorer~\cite{src-solexp} and ClickHouse public
datasets~\cite{src-clickhouse} for Solana, BscScan~\cite{src-bsc} for BNB Smart
Chain, and Tronscan~\cite{src-tron} for TRON. Transactions per block for Ethereum and TRON were computed using Dune
transaction data from 23 May to 20 August 2026 inclusive, excluding blocks
containing no transactions~\cite{dune}; the queries are given in
Appendix~\ref{app:dune}. Solana's
ledger size is the approximate figure reported by Triton One and
Solana~\cite{tritonsuperbank,solana400tb}; BNB Smart Chain's was computed as
cumulative block data from BscScan's daily series.

\section{Experimental Results}
\label{sec:results}
This section presents the experimental results for the 13 signature
configurations. All results are based on the 10,000-transaction workload
described in Section~\ref{sec:workload}, and reported values are medians across
the ten runs of each verification mode unless stated otherwise. SLH-DSA denotes
SLH-DSA-SHA2-128s throughout.

\subsection{Cryptographic Performance}
Table~\ref{tab:crypto-performance} summarises the signing and verification
performance under scheme-level verification. 
Run-to-run variation was small: for every configuration, the interquartile
range of each reported timing and throughput measurement was at most 2.2\% of
the median.
All comparisons refer to the implementations and platform described in Section~\ref{sec:implement}.  As expected, ECDSA provides the
lowest signing cost, with a median of 20.54~$\mu$s per transaction. ML-DSA-44
and Falcon-512 require 185.13~$\mu$s and 205.29~$\mu$s, respectively,
making signing approximately one order of magnitude slower than ECDSA.

ML-DSA-65 and ML-DSA-87 require 251.20 and 304.57~$\mu$s per transaction
for signing, respectively, compared with 185.13~$\mu$s for ML-DSA-44.
Falcon-1024 requires 334.11~$\mu$s, compared with 205.29~$\mu$s for
Falcon-512. The additional parameter sets therefore have higher measured
signing costs than their respective smaller configurations.

The difference is considerably larger for SLH-DSA-SHA2-128s, for which
the median signing time is 279.03~ms per transaction. Its signing cost is
therefore around three orders of magnitude greater than those of ML-DSA-44
and Falcon-512. This is also reflected in crypto-only throughput, which
falls from 18,748~tx/s for ECDSA to 4,325~tx/s for ML-DSA-44 and
4,251~tx/s for Falcon-512, and to 3.58~tx/s for SLH-DSA-SHA2-128s.

Verification exhibits a different pattern. Falcon-512 records a median
verification time of 30.06~$\mu$s per transaction, comparable to the 32.77~$\mu$s measured for ECDSA. ML-DSA-44 requires 46.18~$\mu$s,
while SLH-DSA-SHA2-128s requires 300.08~$\mu$s. Thus, the particularly
high computational cost observed for SLH-DSA is primarily associated
with signing rather than verification.

The median verification times for ML-DSA-65, ML-DSA-87, and Falcon-1024
are 66.00, 106.00, and 53.50~$\mu$s per transaction, respectively.
Although ML-DSA-87 signs faster than Falcon-1024, its slower verification
results in lower crypto-only throughput: 2,435.83 versus 2,582.47~tx/s.

\begin{table}[t]
\centering
\small
\setlength{\tabcolsep}{4pt}
\caption{Median cryptographic performance under scheme-level verification.
For hybrid configurations, the NIST category is that of the post-quantum
component.}
\label{tab:crypto-performance}
\begin{tabular}{|l|c|r|r|r|}
\hline
Configuration & \shortstack{NIST\\cat.} & Sign ($\mu$s/tx) & Verify ($\mu$s/tx) &
\shortstack{Crypto throughput\\(tx/s)} \\
\hline
ECDSA             & ---  & 20.54     & 32.77  & 18,748.35 \\
Falcon-512        & 1    & 205.29    & 30.06  & 4,250.57 \\
SLH-DSA           & 1    & 279,025.87 & 300.08 & 3.58 \\
ML-DSA-44         & 2    & 185.13    & 46.18  & 4,324.59 \\
ML-DSA-65         & 3    & 251.20    & 66.00  & 3,154.94 \\
ML-DSA-87         & 5    & 304.57    & 106.00 & 2,435.83 \\
Falcon-1024       & 5    & 334.11    & 53.50  & 2,582.47 \\
ECDSA+Falcon-512  & 1    & 231.98    & 64.39  & 3,373.68 \\
ECDSA+SLH-DSA     & 1    & 279,084.96 & 336.92 & 3.58 \\
ECDSA+ML-DSA-44   & 2    & 206.57    & 79.48  & 3,497.28 \\
ECDSA+ML-DSA-65   & 3    & 275.24    & 99.98  & 2,664.12 \\
ECDSA+ML-DSA-87   & 5    & 329.00    & 140.46 & 2,129.78 \\
ECDSA+Falcon-1024 & 5    & 359.90    & 88.88  & 2,221.28 \\
\hline
\end{tabular}
\end{table}

The hybrid configurations introduce the expected additional cost of
performing both classical and post-quantum operations. ECDSA+ML-DSA-44
requires 206.57~$\mu$s for signing and 79.48~$\mu$s for verification,
while ECDSA+Falcon-512 requires 231.98~$\mu$s and 64.39~$\mu$s,
respectively. Their crypto-only throughputs fall to 3,497 and
3,374~tx/s. In contrast, adding ECDSA to SLH-DSA has little effect on
overall cryptographic throughput because the execution time remains
dominated by SLH-DSA signing.
For ECDSA+ML-DSA-65, ECDSA+ML-DSA-87, and ECDSA+Falcon-1024,
crypto-only throughput is 2,664.12, 2,129.78, and 2,221.28~tx/s,
respectively. Each is lower than that of its corresponding post-quantum-only
configuration, reflecting the additional classical signing and verification
operations.

\subsection{End-to-End Blockchain Performance}
The differences in cryptographic cost are also visible at the system
level. ECDSA achieves a median end-to-end throughput of 16,008~tx/s.
This decreases to 4,094~tx/s for ML-DSA-44 and 4,030~tx/s for
Falcon-512. Relative to their respective post-quantum-only
configurations, hybridisation reduces throughput further, to
3,325~tx/s for ECDSA+ML-DSA-44 and 3,210~tx/s for ECDSA+Falcon-512.
ML-DSA-65, ML-DSA-87, and Falcon-1024 achieve
3,033, 2,360, and 2,494~tx/s, respectively. Their ECDSA
hybrids achieve 2,563, 2,063, and 2,145~tx/s, respectively.
The throughput reduction associated with hybridisation is therefore also
observed for all three additional post-quantum parameter sets.

The effect is again much larger for SLH-DSA. Both SLH-DSA and
ECDSA+SLH-DSA achieve approximately 3.58~tx/s. The similarity between
these two values indicates that adding the ECDSA operation contributes
little to the total execution time when compared with the much larger
SLH-DSA signing cost.

Moving verification from the transaction-processing stage to block
commitment does not materially change this overall pattern. Median
end-to-end throughput in commit-level mode is 16,028~tx/s for ECDSA,
4,096~tx/s for ML-DSA-44, and 4,042~tx/s for Falcon-512. The
corresponding hybrid configurations achieve 3,377~tx/s and 3,247~tx/s,
respectively. SLH-DSA and ECDSA+SLH-DSA remain at approximately
3.60~tx/s. This suggests that, in the present experimental framework,
the choice of signature scheme has a much greater effect on overall
throughput than the point at which verification is performed.
In commit-level mode, ML-DSA-65, ML-DSA-87, and Falcon-1024 achieve
3,042, 2,378, and 2,518~tx/s, respectively. Their ECDSA
hybrids achieve 2,577, 2,050, and 2,169~tx/s, respectively.
The ordering of these six configurations remains unchanged between the
two verification modes, although the small throughput changes do not
consistently favour one mode.

Under commit-level verification, commitment-time measurements show a
similar relationship to the verification costs. Median commitment time is
33.16~$\mu$s per transaction for ECDSA, 46.69~$\mu$s for ML-DSA-44, and
29.00~$\mu$s for Falcon-512. The hybrid configurations increase this
to 78.36~$\mu$s for ECDSA+ML-DSA-44 and 62.39~$\mu$s for
ECDSA+Falcon-512. SLH-DSA and ECDSA+SLH-DSA require 282.71 and
313.80~$\mu$s per transaction, respectively.
The commit-level commitment times for ML-DSA-65, ML-DSA-87, and
Falcon-1024 are 66.43, 106.59, and 51.61~$\mu$s per transaction,
respectively. Their ECDSA hybrids require
99.38, 142.43, and 85.87~$\mu$s per transaction, respectively.

\subsection{Signature and Storage Overheads}
The signature configurations also differ substantially in their storage
requirements. Because the adjusted sizes exclude public keys, they reflect
signature and other transaction content only; public-key sizes are reported
separately.
Table~\ref{tab:size-results} reports the mean signature and public-key
sizes together with the adjusted transaction sizes and ledger totals (defined in Section~\ref{sec:metrics}) for all 13 configurations.

\begingroup
\raggedbottom
\par\medskip
\noindent\begin{minipage}{\linewidth}
\centering
\small
\setlength{\tabcolsep}{3pt}
\makeatletter
\def\@captype{table}
\makeatother
\caption{Mean signature sizes (over all generated signatures), public-key sizes, mean
adjusted transaction sizes (bytes) and median adjusted ledger sizes (MB;
1~MB $= 10^6$ bytes) under scheme-level verification for the
10,000-transaction workload.}
\label{tab:size-results}
\begin{tabular}{|l|r|r|r|r|}
\hline
Configuration & Signature & Public key  & Adj.\ tx  & Adj.\ ledger (MB) \\
\hline
ECDSA             & 70.49   & 33   & 538   & 6.05   \\
Falcon-512        & 655.05  & 897  & 1,695  & 17.63  \\
SLH-DSA           & 7,856    & 32   & 16,108 & 161.76 \\
ML-DSA-44         & 2,420    & 1,312 & 5,224  & 52.91  \\
ML-DSA-65         & 3,309    & 1,952 & 7,001  & 70.69  \\
ML-DSA-87         & 4,627    & 2,592 & 9,637  & 97.05  \\
Falcon-1024       & 1,270.67 & 1,793 & 2,927  & 29.95  \\
ECDSA+Falcon-512  & 725.54  & 930  & 1,867  & 19.35  \\
ECDSA+SLH-DSA     & 7,926.49  & 65   & 16,280 & 163.48 \\
ECDSA+ML-DSA-44   & 2,490.49 & 1,345 & 5,396  & 54.63  \\
ECDSA+ML-DSA-65   & 3,379.49 & 1,985 & 7,173  & 72.41  \\
ECDSA+ML-DSA-87   & 4,697.49  & 2,625 & 9,809  & 98.77  \\
ECDSA+Falcon-1024 & 1,341.16  & 1,826 & 3,099  & 31.67  \\
\hline
\end{tabular}
\end{minipage}
\par\medskip

Among the post-quantum schemes, Falcon-512 produces the smallest
signatures, with a mean size of 655.05 bytes, compared with 2,420 bytes
for ML-DSA-44 and 7,856 bytes for SLH-DSA. The hybrid configurations
increase these values by approximately the size of the additional ECDSA
signature. ECDSA and Falcon signatures are variable in length, which
accounts for the non-integer mean sizes of these configurations and their
hybrids.

Both signature and public-key sizes increase from ML-DSA-44 to ML-DSA-65
and ML-DSA-87. Falcon-1024 also has larger signatures and public keys
than Falcon-512, but both remain smaller than those of ML-DSA-65 and
ML-DSA-87. The same size ordering holds for their ECDSA hybrids.

The same ordering is visible in the serialised transaction sizes. The mean
adjusted transaction size is 538 bytes for ECDSA, increasing to 1,695 bytes for
Falcon-512, 5,224 bytes for ML-DSA-44, and 16,108 bytes for SLH-DSA. The
corresponding hybrid transaction sizes are 1,867, 5,396, and 16,280 bytes. For
ML-DSA-65, ML-DSA-87, and Falcon-1024, the adjusted transaction sizes are
7,001, 9,637, and 2,927 bytes, respectively. Their ECDSA hybrids require 7,173,
9,809, and 3,099 bytes, respectively.

Over the 10,000-transaction workload, these differences accumulate in the
experimental ledger. The adjusted ledger size is 6.05~MB for ECDSA,
17.63~MB for Falcon-512, 29.95~MB for Falcon-1024, and 52.91, 70.69 and
97.05~MB for ML-DSA-44, ML-DSA-65 and ML-DSA-87, respectively, rising to
161.76~MB for SLH-DSA. The corresponding ECDSA hybrids require 19.35,
31.67, 54.63, 72.41, 98.77 and 163.48~MB.
Falcon-1024 and its hybrid therefore require less ledger storage than any
ML-DSA or SLH-DSA configuration. The ledger representation also stores
per-block metadata (index, timestamp, hash and previous-block hash) and a list
of transaction identifiers. The latter duplicates the identifier held in each
transaction, adding approximately 67~bytes per transaction. These measurements illustrate how increases at the
individual-signature and transaction level accumulate as the number of
stored transactions grows.

The transaction and ledger measurements in this experiment are based on
the JSON representation used by the prototype and should therefore not
be interpreted as the native transaction sizes of deployed blockchain
protocols. The real-world scaling analysis consequently considers
signature-size overhead separately from the prototype-specific
serialisation overhead.
\par
\endgroup

\subsection{Real-World Blockchain Scaling}
\label{sec:real-world-scaling}

\begingroup
\raggedbottom
The following storage projections use the mean signature sizes measured in
our experiments. The ECDSA configuration and the ECDSA component of each
hybrid configuration use the measured mean DER-encoded signature size of
70.49~bytes, which differs from the representative native signature sizes
defined in Section~\ref{sec:realworld}.

\par\medskip
\noindent\begin{minipage}{\linewidth}
\centering
\small
\setlength{\abovecaptionskip}{0pt}
\setlength{\belowcaptionskip}{10pt}
\setlength{\tabcolsep}{2.2pt}
\makeatletter
\def\@captype{table}
\makeatother
\caption{Annual total signature storage (TB/year).}
\label{tab:scaling-annual-signature-storage}
\par
\begin{tabular}{|l|r|r|r|r|r|}
\hline
Configuration & Bitcoin & Ethereum & Solana & \shortstack{BNB Smart\\Chain} & TRON \\
\hline
ECDSA             & 0.017 & 0.058 & 7.305 & 0.405 & 0.312 \\
Falcon-512 & 0.160 & 0.537 & 67.877 & 3.762 & 2.899 \\
SLH-DSA           & 1.916 & 6.440 & 814.043 & 45.118 & 34.773 \\
ML-DSA-44         & 0.590 & 1.984 & 250.762 & 13.898 & 10.712 \\
ML-DSA-65         & 0.807 & 2.713 & 342.881 & 19.004 & 14.647 \\
ML-DSA-87         & 1.129 & 3.793 & 479.452 & 26.574 & 20.480 \\
Falcon-1024 & 0.310 & 1.042 & 131.668 & 7.298 & 5.624 \\
ECDSA+Falcon-512 & 0.177 & 0.595 & 75.181 & 4.167 & 3.211 \\
ECDSA+SLH-DSA & 1.933 & 6.498 & 821.348 & 45.523 & 35.085 \\
ECDSA+ML-DSA-44 & 0.607 & 2.042 & 258.066 & 14.303 & 11.024 \\
ECDSA+ML-DSA-65 & 0.824 & 2.770 & 350.185 & 19.409 & 14.959 \\
ECDSA+ML-DSA-87 & 1.146 & 3.851 & 486.757 & 26.978 & 20.793 \\
ECDSA+Falcon-1024 & 0.327 & 1.099 & 138.972 & 7.702 & 5.936 \\
\hline
\end{tabular}
\end{minipage}
\par\medskip
\par\medskip
\noindent\begin{minipage}{\linewidth}
\centering
\small
\setlength{\abovecaptionskip}{0pt}
\setlength{\belowcaptionskip}{10pt}
\setlength{\tabcolsep}{2.2pt}
\makeatletter
\def\@captype{table}
\makeatother
\caption{Lifetime total signature storage (TB).}
\label{tab:scaling-lifetime-signature-storage}
\par

\begin{tabular}{|l|r|r|r|r|r|}
\hline
Configuration & Bitcoin & Ethereum & Solana & \shortstack{BNB Smart\\Chain} & TRON \\
\hline
ECDSA & 0.100 & 0.258 & 37.864 & 1.010 & 1.064 \\

Falcon-512 & 0.925 & 2.401 & 351.837 & 9.387 & 9.891 \\

SLH-DSA           & 11.092 & 28.795 & 4,219.574 & 112.574 & 118.618 \\
ML-DSA-44         & 3.417 & 8.870 & 1,299.818 & 34.678 & 36.540 \\
ML-DSA-65         & 4.672 & 12.129 & 1,777.313 & 47.417 & 49.962 \\
ML-DSA-87         & 6.533 & 16.960 & 2,485.230 & 66.303 & 69.863 \\

Falcon-1024 & 1.794 & 4.657 & 682.498 & 18.208 & 19.186 \\

ECDSA+Falcon-512 & 1.024 & 2.659 & 389.701 & 10.397 & 10.955 \\

ECDSA+SLH-DSA & 11.192 & 29.053 & 4,257.437 & 113.584 & 119.682 \\

ECDSA+ML-DSA-44 & 3.516 & 9.129 & 1,337.681 & 35.688 & 37.604 \\

ECDSA+ML-DSA-65 & 4.772 & 12.387 & 1,815.177 & 48.427 & 51.027 \\

ECDSA+ML-DSA-87 & 6.632 & 17.218 & 2,523.094 & 67.313 & 70.927 \\

ECDSA+Falcon-1024 & 1.894 & 4.916 & 720.358 & 19.218 & 20.250 \\
\hline
\end{tabular}
\end{minipage}
\par\medskip

Solana dominates absolute storage because it accounts for 90.2\% of combined daily
transactions and 94.0\% of lifetime transactions across the five networks. BNB and TRON
illustrate a different distinction: BNB has 29.8\% higher daily transaction volume, but
TRON has 5.4\% more historical transactions. Consequently, BNB has higher annual storage
requirements, while TRON has higher lifetime totals.

Moving to higher parameter sets within each signature family increases signature storage across all five networks.
ML-DSA-65 requires 36.7\% more signature storage than ML-DSA-44, while ML-DSA-87
requires 91.2\% more. Falcon-1024 requires 94.0\% more than Falcon-512. Choosing
Falcon-1024 instead of ML-DSA-87 at Solana's volume reduces modelled signature storage by 347.78~TB/year and 1,802.73~TB across its recorded history.

Adding ECDSA increases signature storage by 10.76\% for Falcon-512, 5.55\% for
Falcon-1024, 2.91\% for ML-DSA-44, 2.13\% for ML-DSA-65, 1.52\% for ML-DSA-87
and 0.90\% for SLH-DSA. Although these percentages differ, all six pairs add
approximately 70.5~bytes per transaction, the size of the ECDSA signature. That
becomes roughly 17~GB/year on Bitcoin, 58~GB on Ethereum, 0.40~TB on BNB,
0.31~TB on TRON and 7.3~TB on Solana. Across Solana's historical transactions,
retaining ECDSA adds approximately 37.9~TB.

The following table presents the projected total ledger sizes, in TB, for each signature
configuration, holding the non-signature component fixed.

\par\medskip
\noindent\begin{minipage}{\linewidth}
\centering
\small
\setlength{\abovecaptionskip}{0pt}
\setlength{\belowcaptionskip}{10pt}
\setlength{\tabcolsep}{2.2pt}
\makeatletter
\def\@captype{table}
\makeatother
\caption{
Counterfactual historical ledger sizes (TB), with the non-signature component held fixed.
}
\label{tab:scaling-projected-ledger-size}
\par
\begin{tabular}{|l|r|r|r|r|r|}
\hline
Configuration & Bitcoin & Ethereum & Solana & \shortstack{BNB Smart\\Chain} & TRON \\
\hline
ECDSA & 0.762 & 1.765 & 403.488 & 7.095 & 3.980 \\

Falcon-512 & 1.587 & 3.908 & 717.462 & 15.471 & 12.806 \\

SLH-DSA           & 11.754 & 30.301 & 4,585.199 & 118.658 & 121.533 \\
ML-DSA-44         & 4.079 & 10.377 & 1,665.443 & 40.762 & 39.455 \\
ML-DSA-65         & 5.335 & 13.635 & 2,142.938 & 53.501 & 52.878 \\
ML-DSA-87         & 7.195 & 18.466 & 2,850.855 & 72.388 & 72.778 \\

Falcon-1024 & 2.457 & 6.164 & 1,048.122 & 24.293 & 22.101 \\

ECDSA+Falcon-512 & 1.687 & 4.166 & 755.325 & 16.481 & 13.870 \\

ECDSA+SLH-DSA & 11.854 & 30.560 & 4,623.062 & 119.668 & 122.597 \\

ECDSA+ML-DSA-44 & 4.179 & 10.635 & 1,703.306 & 41.772 & 40.519 \\

ECDSA+ML-DSA-65 & 5.434 & 13.894 & 2,180.801 & 54.512 & 53.942 \\

ECDSA+ML-DSA-87 & 7.295 & 18.724 & 2,888.719 & 73.398 & 73.843 \\

ECDSA+Falcon-1024 & 2.556 & 6.422 & 1,085.982 & 25.303 & 23.166 \\
\hline
\end{tabular}
\end{minipage}
\par\medskip

\par\medskip
\noindent\begin{minipage}{\linewidth}
\centering
\small
\setlength{\abovecaptionskip}{0pt}
\setlength{\belowcaptionskip}{10pt}
\setlength{\tabcolsep}{2.2pt}
\makeatletter
\def\@captype{table}
\makeatother
\caption{Percentage changes relative to each existing ledger.}
\label{tab:scaling-ledger-percentage-change}
\par

\begin{tabular}{|l|r|r|r|r|r|}
\hline
Configuration & Bitcoin & Ethereum & Solana & \shortstack{BNB Smart\\Chain} & TRON \\
\hline
ECDSA & $-$0.09\% & 1.15\% & 0.87\% & 1.12\% & 2.13\% \\

Falcon-512 & 108.12\% & 123.95\% & 79.37\% & 120.51\% & 228.62\% \\

SLH-DSA           & 1,441.12\% & 1,636.69\% & 1,046.30\% & 1,591.27\% & 3,018.71\% \\

ML-DSA-44         & 434.84\% & 494.73\% & 316.36\% & 481.00\% & 912.47\% \\

ML-DSA-65         & 599.40\% & 681.48\% & 435.73\% & 662.57\% & 1,256.92\% \\

ML-DSA-87         & 843.39\% & 958.36\% & 612.71\% & 931.76\% & 1,767.60\% \\

Falcon-1024 & 222.08\% & 253.28\% & 162.03\% & 246.25\% & 467.15\% \\

ECDSA+Falcon-512 & 121.17\% & 138.76\% & 88.83\% & 134.91\% & 255.93\% \\

ECDSA+SLH-DSA     & 1,454.17\% & 1,651.50\% & 1,055.77\% & 1,605.67\% & 3,046.02\% \\

ECDSA+ML-DSA-44 & 447.89\% & 509.53\% & 325.83\% & 495.39\% & 939.78\% \\

ECDSA+ML-DSA-65 & 612.45\% & 696.29\% & 445.20\% & 676.97\% & 1,284.24\% \\

ECDSA+ML-DSA-87   & 856.44\% & 973.17\% & 622.18\% & 946.16\% & 1,794.91\% \\

ECDSA+Falcon-1024 & 235.13\% & 268.09\% & 171.50\% & 260.65\% & 494.46\% \\
\hline
\end{tabular}
\end{minipage}
\par\medskip

With Falcon-512, Bitcoin's additional storage above its existing signature baseline is
0.390 GB/day, 0.142 TB/year and 0.825 TB over its recorded history. Ethereum's
corresponding additions are 1.325 GB/day, 0.484 TB/year and 2.163 TB. Ethereum
has 3.36 times Bitcoin's daily traffic, but only 2.60 times its historical transactions
and an existing ledger 2.29 times as large. Their percentage ledger increases are
therefore much closer than their daily storage additions.

   Under the selected storage baselines, percentage growth follows the same order
   for every PQ and hybrid configuration: TRON, Ethereum, BNB, Bitcoin, then
   Solana. These percentages are relative to baselines that differ in scope across
   chains (Section~\ref{sec:realworld}); close values, such as those for Ethereum and BNB Smart
   Chain, should therefore not be read as a general ranking of the networks.
   Adding raw signature bytes to these baselines also assumes an approximately
   one-for-one increase in stored data, without accounting for compression or
   database overhead.
This differs from the
absolute-storage ranking. Falcon-512 adds 317.46 TB to Solana's modelled historical
storage, compared with 8.455 TB on BNB, yet the percentage increases are 79.37\% and
120.51\%, respectively. Solana's much larger starting ledger explains the reversal. Its 400 TB baseline is approximate, and BNB’s baseline counts block data only, so the denominators also differ somewhat in scope.

The same distinction applies when ECDSA is added to a PQ signature. Retaining ECDSA adds
roughly 9.5 percentage points to Solana's projected ledger growth and 14.4 points to
BNB's, despite the much larger absolute storage addition on Solana.

The small ledger changes shown for ECDSA reflect differences between its measured mean signature size of 70.49~bytes and the representative native
signature sizes of 64 to 71~bytes, rather than the overhead of introducing  post-quantum signatures.

The next three tables report estimated serial processing hours for each day's
transaction activity, extrapolated from the experimental measurements. These are
benchmark-equivalent workloads, not actual production-network execution times.

\par\medskip
\noindent\begin{minipage}{\linewidth}
\centering
\small
\setlength{\abovecaptionskip}{0pt}
\setlength{\belowcaptionskip}{10pt}
\setlength{\tabcolsep}{2.2pt}
\makeatletter
\def\@captype{table}
\makeatother
\caption{Daily end-to-end processing workload (hours/day).}
\label{tab:scaling-daily-end-to-end-workload}
\par
\begin{tabular}{|l|r|r|r|r|r|}
\hline
Configuration & Bitcoin & Ethereum & Solana & \shortstack{BNB Smart\\Chain} & TRON \\
\hline
ECDSA             & 0.012 & 0.039 & 4.93 & 0.273 & 0.210 \\
Falcon-512        & 0.046 & 0.155 & 19.57 & 1.08 & 0.836 \\

SLH-DSA & 51.86 & 174.29 & 22,030.60 & 1,221.04 & 941.07 \\

ML-DSA-44         & 0.045 & 0.152 & 19.26 & 1.07 & 0.823 \\

ML-DSA-65         & 0.061 & 0.206 & 26.00 & 1.44 & 1.11 \\

ML-DSA-87         & 0.079 & 0.264 & 33.41 & 1.85 & 1.43 \\

Falcon-1024       & 0.074 & 0.250 & 31.62 & 1.75 & 1.35 \\

ECDSA+Falcon-512  & 0.058 & 0.194 & 24.56 & 1.36 & 1.05 \\

ECDSA+SLH-DSA & 51.88 & 174.35 & 22,038.46 & 1,221.48 & 941.40 \\

ECDSA+ML-DSA-44   & 0.056 & 0.188 & 23.72 & 1.31 & 1.01 \\

ECDSA+ML-DSA-65   & 0.072 & 0.243 & 30.76 & 1.70 & 1.31 \\

ECDSA+ML-DSA-87   & 0.090 & 0.302 & 38.22 & 2.12 & 1.63 \\

ECDSA+Falcon-1024 & 0.087 & 0.291 & 36.76 & 2.04 & 1.57 \\
\hline
\end{tabular}
\end{minipage}
\par\medskip

\par\medskip
\noindent\begin{minipage}{\linewidth}
\centering
\small
\setlength{\abovecaptionskip}{0pt}
\setlength{\belowcaptionskip}{10pt}
\setlength{\tabcolsep}{2.2pt}
\makeatletter
\def\@captype{table}
\makeatother
\caption{Daily signing workload (hours/day).}
\label{tab:scaling-daily-signing-workload}
\par
\begin{tabular}{|l|r|r|r|r|r|}
\hline
Configuration & Bitcoin & Ethereum & Solana & \shortstack{BNB Smart\\Chain} & TRON \\
\hline
ECDSA             & 0.0038 & 0.0128 & 1.62 & 0.0898 & 0.0692 \\
Falcon-512        & 0.0381 & 0.1281 & 16.19 & 0.8973 & 0.6915 \\
SLH-DSA & 51.80 & 174.08 & 22,003.67 & 1,219.55 & 939.92 \\
ML-DSA-44         & 0.0344 & 0.1155 & 14.60 & 0.8091 & 0.6236 \\
ML-DSA-65         & 0.0466 & 0.1567 & 19.81 & 1.10 & 0.8462 \\
ML-DSA-87         & 0.0565 & 0.1900 & 24.02 & 1.33 & 1.03 \\
Falcon-1024       & 0.0620 & 0.2084 & 26.35 & 1.46 & 1.13 \\
ECDSA+Falcon-512  & 0.0431 & 0.1447 & 18.29 & 1.01 & 0.7814 \\
ECDSA+SLH-DSA & 51.81 & 174.11 & 22,008.33 & 1,219.81 & 940.12 \\
ECDSA+ML-DSA-44   & 0.0383 & 0.1289 & 16.29 & 0.9028 & 0.6958 \\
ECDSA+ML-DSA-65   & 0.0511 & 0.1717 & 21.71 & 1.20 & 0.9272 \\
ECDSA+ML-DSA-87   & 0.0611 & 0.2053 & 25.94 & 1.44 & 1.11 \\
ECDSA+Falcon-1024 & 0.0668 & 0.2245 & 28.38 & 1.57 & 1.21 \\
\hline
\end{tabular}
\end{minipage}
\par\medskip

\par\medskip
\noindent\begin{minipage}{\linewidth}
\centering
\small
\setlength{\abovecaptionskip}{0pt}
\setlength{\belowcaptionskip}{10pt}
\setlength{\tabcolsep}{2.2pt}
\makeatletter
\def\@captype{table}
\makeatother
\caption{Daily verification workload (hours/day).}
\label{tab:scaling-daily-verification-workload}
\par
\begin{tabular}{|l|r|r|r|r|r|}
\hline
Configuration & Bitcoin & Ethereum & Solana & \shortstack{BNB Smart\\Chain} & TRON \\
\hline
ECDSA             & 0.0061 & 0.0204 & 2.58 & 0.1432 & 0.1104 \\
Falcon-512        & 0.0056 & 0.0188 & 2.37 & 0.1314 & 0.1013 \\
SLH-DSA & 0.0557 & 0.1872 & 23.66 & 1.31 & 1.01 \\
ML-DSA-44         & 0.0086 & 0.0288 & 3.64 & 0.2018 & 0.1556 \\
ML-DSA-65         & 0.0123 & 0.0412 & 5.20 & 0.2885 & 0.2223 \\
ML-DSA-87         & 0.0197 & 0.0661 & 8.36 & 0.4633 & 0.3571 \\
Falcon-1024       & 0.0099 & 0.0334 & 4.22 & 0.2339 & 0.1802 \\
ECDSA+Falcon-512  & 0.0120 & 0.0402 & 5.08 & 0.2814 & 0.2169 \\
ECDSA+SLH-DSA & 0.0625 & 0.2102 & 26.57 & 1.47 & 1.13 \\
ECDSA+ML-DSA-44   & 0.0148 & 0.0496 & 6.27 & 0.3474 & 0.2677 \\
ECDSA+ML-DSA-65   & 0.0186 & 0.0624 & 7.88 & 0.4370 & 0.3368 \\
ECDSA+ML-DSA-87   & 0.0261 & 0.0876 & 11.08 & 0.6139 & 0.4732 \\
ECDSA+Falcon-1024 & 0.0165 & 0.0554 & 7.01 & 0.3885 & 0.2994 \\
\hline
\end{tabular}
\end{minipage}
\par\medskip

All ML-DSA and Falcon configurations are estimated to require at most 2.12 hours to
process one day's transactions on the four lower-volume blockchains. On Solana, the six
additional configurations (ML-DSA-65, ML-DSA-87, Falcon-1024 and their hybrids) are
estimated to require between 26.00 and 38.22 hours. These estimates are based on our
experimental results, not actual blockchain processing times.

The component tables reveal differences hidden by the end-to-end totals. On Solana,
Falcon-512 requires more signing work than ML-DSA-44, 16.19 versus 14.60
hours/day, but less verification work, 2.37 versus 3.64 hours/day. The
higher-parameter comparison shows the same reversal: Falcon-1024 requires more signing
than ML-DSA-87, 26.35 versus 24.02 hours, but approximately half the
verification workload, 4.22 versus 8.36 hours. A configuration's overall
processing estimate therefore cannot substitute for its verification-only estimate.

Adding ECDSA to a PQ signature increases verification work more sharply than signature
storage. On Solana, the Falcon-512 hybrid raises verification from 2.37 to 5.08
hours/day, while increasing signature storage by 10.76\%. For ML-DSA-87, verification
rises from 8.36 to 11.08 hours/day, while signature storage increases by only
1.52\%. Across the five ML-DSA and Falcon pairs, retaining ECDSA adds approximately
1.69 to 2.10 signing hours and 2.63 to 2.79 verification hours/day at Solana's
volume.

SLH-DSA shows the largest separation between signing and verification. Bitcoin's modelled
workload is 51.80 signing hours/day but only 0.0557 verification hours/day. On
Solana, the corresponding figures are 22,003.67 and 23.66 hours/day. Its
enormous end-to-end estimates are therefore largely associated with signing. The signing
and verification projections are calculated from separate benchmark inputs and should
not be treated as an exact accounting breakdown of the end-to-end table.

The final table shows the total signature payload per representative block or slot, in
MB, calculated using each network's recorded average transactions per block or slot.
These estimates exclude all other block contents.

\par\medskip
\noindent\begin{minipage}{\linewidth}
\centering
\small
\setlength{\abovecaptionskip}{0pt}
\setlength{\belowcaptionskip}{10pt}
\setlength{\tabcolsep}{2.2pt}
\makeatletter
\def\@captype{table}
\makeatother
\caption{Total signature payload per representative block or slot (MB).}
\label{tab:scaling-block-signature-payload}
\par
\begin{tabular}{|l|r|r|r|r|r|}
\hline
Configuration & Bitcoin & Ethereum & Solana & \shortstack{BNB Smart\\Chain} & TRON \\
\hline
ECDSA & 0.3240 & 0.0221 & 0.0947 & 0.0058 & 0.0297 \\

Falcon-512 & 3.011 & 0.2050 & 0.8803 & 0.0537 & 0.2759 \\

SLH-DSA           & 36.112 & 2.459 & 10.558 & 0.6443 & 3.308 \\
ML-DSA-44         & 11.124 & 0.7574 & 3.252 & 0.1985 & 1.019 \\
ML-DSA-65         & 15.211 & 1.036 & 4.447 & 0.2714 & 1.394 \\
ML-DSA-87         & 21.269 & 1.448 & 6.218 & 0.3795 & 1.949 \\

Falcon-1024 & 5.841 & 0.3977 & 1.708 & 0.1042 & 0.5351 \\

ECDSA+Falcon-512 & 3.335 & 0.2271 & 0.9751 & 0.0595 & 0.3055 \\

ECDSA+SLH-DSA & 36.436 & 2.481 & 10.652 & 0.6501 & 3.338 \\

ECDSA+ML-DSA-44 & 11.448 & 0.7794 & 3.347 & 0.2042 & 1.049 \\

ECDSA+ML-DSA-65 & 15.535 & 1.058 & 4.542 & 0.2772 & 1.423 \\

ECDSA+ML-DSA-87   & 21.593 & 1.470 & 6.313 & 0.3852 & 1.978 \\

ECDSA+Falcon-1024 & 6.165 & 0.4197 & 1.802 & 0.1100 & 0.5648 \\
\hline
\end{tabular}
\end{minipage}
\par\medskip

Bitcoin has the largest representative per-block signature payload for every
configuration, despite having the lowest daily transaction volume. Its representative
block contains 4,596.73 transactions, compared with 1,343.90 for Solana and 82.01 for
BNB. Consequently, Bitcoin's projected signature payload per block is 3.42 times
Solana's, even though Solana's annual storage requirement is much greater. The
per-block ordering is Bitcoin, Solana, TRON, Ethereum and BNB.

The same reversal affects hybrid overhead. Retaining ECDSA adds approximately
0.324~MB per representative Bitcoin block, compared with about
0.0058 MB for BNB. BNB nevertheless accumulates much more additional hybrid
storage annually. Per-block payload and annual accumulation therefore describe different
parts of the scaling cost.

Even with Falcon-512, the smallest evaluated post-quantum signature, the projected signature payload per block exceeds the recorded average total block size on Bitcoin, Ethereum and TRON, and reaches 98\% of it on BNB Smart Chain. Every other evaluated PQ and hybrid configuration exceeds the average block size on all four of these chains. These comparisons use average block
sizes rather than protocol maximums.
\par
\endgroup

\subsection{Summary of Experimental Findings}

The results show that post-quantum migration involves different
computational and storage trade-offs depending on the signature scheme.
ML-DSA-44 and Falcon-512 both reduce transaction-processing throughput
relative to ECDSA, although Falcon-512 has substantially smaller
signatures and faster verification in the implementation evaluated here.
ML-DSA-44 provides somewhat higher end-to-end throughput than Falcon-512 under scheme-level verification on this platform, but at a considerably higher storage
cost.

Across ML-DSA-44, ML-DSA-65, and ML-DSA-87, the measured signing and
verification costs increase, throughput decreases, and signatures and public
keys become larger. Their ECDSA hybrids follow the same pattern.
Falcon-1024 similarly has higher computational costs and larger
cryptographic objects than Falcon-512, including in hybrid form. The
cross-family ranking depends on both operations: ML-DSA-87 signs faster
than Falcon-1024, but Falcon-1024 verifies faster and achieves higher
end-to-end throughput. This relationship also holds for their ECDSA
hybrids.

SLH-DSA-SHA2-128s presents a markedly different profile. Its public key
is small, but this is accompanied by much larger signatures and,
particularly, substantially higher signing cost. Hybrid configurations
add further computational and storage overhead for ML-DSA and Falcon.
For SLH-DSA, however, the additional ECDSA operation has little visible
effect on throughput because SLH-DSA signing dominates the overall
execution time.
\section{Discussion}
\label{sec:discussion}
Our results suggest that, for nodes responsible for validating transactions and
storing blockchain data, the storage overhead of post-quantum signatures may be
a greater concern than their verification cost. Verification was relatively
inexpensive for all the lattice-based schemes we evaluated. Even at Solana's
transaction volume, the most demanding of these configurations, ECDSA+ML-DSA-87,
would require approximately 11 single-core hours of verification per day under
our benchmark assumptions (Table~\ref{tab:scaling-daily-verification-workload}). This is
equivalent to an average utilisation of less than half a CPU core.
This is a daily average; it does not account for bursts, block-validation deadlines, multiple signatures per transaction or other validation work.
On the other
four chains, every lattice-based configuration requires well under an hour per
day. Signing is considerably more expensive, particularly for SLH-DSA, but
transaction signing is generally performed by users' wallets rather than by
validating nodes. Storage is another matter. 
Had the recorded transaction histories used Falcon-512, which produces the
smallest post-quantum signatures among the schemes we tested, the estimated
ledger sizes would have been between 79\% (Solana) and 229\% (TRON) larger
(Table~\ref{tab:scaling-ledger-percentage-change}), with the other standalone post-quantum schemes resulting in
substantially greater storage overheads.

Hybrid configurations involve a trade-off of their own. Retaining ECDSA
alongside a post-quantum signature adds approximately 70.5~bytes per transaction,
which is small compared with the size of the post-quantum signature itself.
The effect on verification is more noticeable. Since ECDSA verification takes
approximately as long as Falcon-512 verification, the ECDSA+Falcon-512 hybrid
takes more than twice as long to verify as Falcon-512 alone (64.39 versus
30.06~$\mu$s). Whether this additional cost is justified depends partly on
the confidence placed in the newer post-quantum schemes during the transition.
Hybrid deployment also requires careful consideration of the verification rules.
In our construction, the classical and post-quantum schemes independently sign
the same transaction data, and both signatures must verify successfully.
A practical deployment would need to enforce this requirement at the protocol
level to prevent signature-stripping attacks, in which one of the two signatures is omitted or ignored.

Our signature-only projections also exclude potentially substantial public-key storage costs. In their conventional ECDSA-based transaction authentication,
Ethereum, BNB Smart Chain and TRON can recover public keys from signatures
rather than requiring them to be included explicitly. Standard ML-DSA does not provide public-key recovery. The Falcon
specification outlines a key-recovery mode, without proposing it as
part of the specification, in which the signature roughly doubles in
size but the public key is replaced by a short hash. According to
the specification, this reduces the combined public-key and signature
size by about 15\%~\cite{falcon}. This mode is not supported by the
Falcon implementations evaluated in this study. If transactions also had
to carry post-quantum public keys, the storage overhead would increase
considerably. For example, Falcon-512 has an 897-byte public key, which is larger
than its signature. Consistent with the one-signature-per-transaction
model used throughout, we make the simplifying assumption that each
Bitcoin transaction carries one Falcon-512 public key in place of an
existing 33-byte public key. Under this assumption, the projected ledger growth would increase from approximately 108\% to 268\%. This suggests that public-key
management is almost as important as signature size when considering migration
costs. Approaches that register public keys once and reference them in
subsequent transactions~\cite{Wang2026}, or that support public-key recovery
for Falcon~\cite{Marchsreiter2025}, could therefore make a substantial
difference. Including public keys would not, however, change the relative ordering of the
evaluated standalone post-quantum schemes by combined signature and public-key
size: the totals increase from approximately 1,552~bytes for Falcon-512 to
3,064 for Falcon-1024, 3,732 for ML-DSA-44, 5,261 for ML-DSA-65, 7,219 for
ML-DSA-87 and 7,888 for SLH-DSA.

Beyond storage, existing protocol constraints mean that post-quantum signatures
cannot simply replace classical signatures without further changes.    Solana limits legacy and v0 transactions to 1,232~bytes, which is smaller than
   any ML-DSA signature and than Falcon-512 together with its public key. Since
   September 2026, a v1 transaction format has allowed up to
   4,096~bytes~\cite{solanav1}. Falcon-512's signature and public key
   (1,552~bytes) would fit within this budget with room to spare, whereas
   ML-DSA-44's (3,732~bytes) would leave only 364~bytes for all other transaction
   content; ML-DSA-65's (5,261~bytes) would exceed it, and  ML-DSA-87 and SLH-DSA signatures alone exceed it. Fitting within
   the size limit is also separate from native transaction authentication with
   these schemes, which would require protocol support. Bitcoin's existing scripting rules also impose restrictions,
including a 520-byte limit on individual script stack elements, which also applies to Tapscript~\cite{bip342}.
Accommodating the evaluated post-quantum signatures would therefore require
changes to how signatures are represented and validated, potentially through
a soft fork. 
The per-block estimates in Table~\ref{tab:scaling-block-signature-payload} illustrate the scale of the problem: at the observed transaction
volumes, even with Falcon-512, the projected signature data alone
would exceed the recorded average block size on Bitcoin, Ethereum
and TRON, and nearly fill it on BNB Smart Chain.
These constraints show that migration is not
simply a matter of choosing a different signature algorithm.

These findings should be considered alongside the limitations of our
experimental setup. The measurements were obtained using a single-threaded
Python prototype running on one Windows machine, without modelling consensus,
networking or mempool behaviour. Furthermore, ML-DSA was evaluated without
its dedicated optimised AVX2 implementation, which may affect its performance
relative to the other schemes. Absolute execution times in production
blockchain clients may therefore differ from those reported here. The scaling
analysis also assumes one signature per transaction, which can underestimate
the storage requirements of Bitcoin transactions involving multiple inputs.
In addition, the projections use average rather than maximum block sizes and
rely on an approximate ledger-size estimate for Solana. The evaluated schemes
also target different security levels, so differences in performance and
storage requirements should be considered alongside their respective security
guarantees. Our projections are therefore intended to illustrate the relative
scale of the costs involved rather than provide precise forecasts of production
deployment.

Overall, the results indicate that while the verification costs of lattice-based
post-quantum signatures appear manageable under the workloads considered,
their storage requirements and compatibility with existing blockchain protocols
present more substantial challenges. The choice of signature scheme is therefore
only one part of the migration problem. How public keys are managed, how
transactions are represented, and what changes existing protocols can accommodate
will also play an important role in determining the practicality of post-quantum
migration.
\section{Conclusion}
\label{sec:summary}
This paper evaluated 13 classical, post-quantum and hybrid signature
configurations within a common simulated blockchain framework and projected
the measured overheads onto five public blockchains. The results suggest
that, for nodes responsible for transaction validation and ledger storage,
the storage overhead of post-quantum signatures may be a greater challenge
than their verification cost. Verification remained relatively inexpensive for the lattice-based
schemes evaluated, while, under our modelling assumptions, the Bitcoin
ledger would have been roughly twice its recorded size had its
transaction history used Falcon-512. TRON exhibits the largest
proportional increase in projected ledger size, reflecting its low
ratio of existing ledger size to lifetime transaction count.
Hybrid configurations introduce additional verification and storage costs,
while existing transaction and scripting constraints mean that adopting
post-quantum signatures would also require changes to blockchain protocols.

Among the post-quantum schemes evaluated, Falcon-512 produced the smallest signatures and the fastest verification, although Falcon's
floating-point signing introduces additional implementation and security
considerations. Our findings also highlight the importance of public-key
management, since repeatedly storing post-quantum public keys could
substantially increase the cost of migration. Future work could integrate
these schemes into production blockchain clients, evaluate performance on
multi-core and ARM platforms, consider transactions with multiple inputs,
and investigate approaches for reducing signature and public-key storage
overheads.

\section*{Code and Data Availability}
The experimental code and per-run benchmark data are available from the
authors upon reasonable request. The public blockchain statistics used in
the scaling analysis were obtained from the sources cited in
Section~\ref{sec:realworld}, and the transactions-per-block queries for
Ethereum and TRON are given in Appendix~\ref{app:dune}.

\par
\begingroup
\raggedright
\bibliographystyle{splncs04}
\bibliography{references}
\endgroup
\appendix
\section{Transactions-per-Block Queries}
\label{app:dune}
The 90-day average transactions per block for Ethereum and TRON
(Table~\ref{tab:chain-activity}) were computed using the following SQL queries on
Dune Analytics. The queries exclude blocks containing no transactions.

\medskip
\noindent\begin{minipage}{\linewidth}
\textbf{Ethereum.}
\small
\begin{verbatim}
SELECT AVG(tx_count) AS avg_tx_per_block
FROM (
  SELECT block_number, COUNT(*) AS tx_count
  FROM ethereum.transactions
  WHERE block_time >= TIMESTAMP '2026-05-23 00:00:00'
    AND block_time <  TIMESTAMP '2026-08-21 00:00:00'
  GROUP BY block_number
) t
\end{verbatim}
\end{minipage}

\medskip
\noindent\begin{minipage}{\linewidth}
\textbf{TRON.}
\small
\begin{verbatim}
SELECT AVG(tx_count) AS avg_tx_per_block
FROM (
  SELECT block_number, COUNT(*) AS tx_count
  FROM tron.transactions
  WHERE block_time >= TIMESTAMP '2026-05-23 00:00:00'
    AND block_time <  TIMESTAMP '2026-08-21 00:00:00'
  GROUP BY block_number
) t
\end{verbatim}
\end{minipage}

\end{document}